\documentclass{article}

\usepackage{emulateapj,epsf}
\tighten
\newcommand{\kms}{$\,{\rm km\,s^{\scriptscriptstyle -1}}$}
\newcommand{\gtsim}{\ {\raise-0.5ex\hbox{$\buildrel>\over\sim$}}\
}
\newcommand{\ltsim}{\ {\raise-0.5ex\hbox{$\buildrel<\over\sim$}}\
}
\def \etal{{\it et~al.~}}

\def\simlt{\lower.5ex\hbox{$\; \buildrel < \over \sim \;$}}
\def\simgt{\lower.5ex\hbox{$\; \buildrel > \over \sim \;$}}

\slugcomment{}

\begin{document}

\title{{Measuring the Cosmic Equation of State with Counts of Galaxies}}

\author{Jeffrey A. Newman and Marc Davis\altaffilmark{1}} 
\begin{center}
{Department of Astronomy, University of California, Berkeley, CA 94720-3411;}\nl 
{jnewman@astro.berkeley.edu, marc@astro.berkeley.edu}
\end{center}
\altaffiltext{1}{Also Department of Physics, U.C. Berkeley}
\vskip -12pt

\begin{abstract}
The classical d$N/$d$z$ test allows the determination of fundamental
cosmological parameters from the evolution of the cosmic volume
element.  This test is applied by measuring the redshift distribution
of a tracer whose evolution in number density is known.  In the past,
ordinary galaxies have been used as such a tracer; however, in the
absence of a complete theory of galaxy formation, that method is
fraught with difficulties.  In this paper, we propose studying instead
the evolution of the apparent numbers of dark matter halos as a
function of their circular velocity, observable via the linewidths or
rotation speeds of visible galaxies.  Upcoming redshift surveys will
allow the linewidth distribution of galaxies to be determined at both
$z\sim 1$ and the present day.  In the course of studying this test,
we have devised a rapid, improved semi-analytic method for calculating
the circular velocity distribution of dark halos based upon the
analytic mass function of Sheth \etal (1999) and the formation time
distribution of Lacey \& Cole (1993).  We find that if selection
effects are well-controlled and minimal external constraints are
applied, the planned DEEP Redshift Survey could allow the measurement
of the cosmic equation-of-state parameter $w$ to $\pm 10\%$ (as little
as 3\% if $\Omega_m$ has been well-determined from other
observations). This type of test has the potential also to provide a
constraint on any evolution of $w$ such as that predicted by
``tracker'' models.
\end{abstract}

\section{Introduction}

For almost as long as they have been known to be located outside the
Milky Way, galaxies have been used to probe the geometry of the
universe (\cite{hubble26}).  Fairly early in the history of relativistic
cosmology, it was realized that if the number density of galaxies is
known, their apparent numbers per solid angle and redshift interval
depend directly on basic cosmological parameters (\cite{tolman34}).
Applying this ``d$N/$d$z$ test'' requires a substantial sample of
distant galaxies with measured redshifts; it was first applied by Loh
and Spillar (1986) using a set of 406 galaxies with estimated
photometric redshifts $z_p$ between 0.15 and 0.85.  Assuming that the
comoving number density of galaxies is constant and that their
luminosity function retains similar shape over that
redshift range, they measured $\Omega=0.9^{+0.6}_{-0.5}$.

However, it remains unclear whether the assumptions of Loh and Spillar
hold; no complete theory or simulation of galaxy formation and
evolution as yet exists (\cite{benson99}).  It would thus be preferable
to apply such a test to objects whose abundance
can be calculated semi-analytically or via computer models,
eliminating such ambiguities.  In this paper, we suggest that the
observed numbers of galaxies (and thus indirectly dark halos) as a
function of linewidth and redshift provide a candidate for improving
on the analysis of Loh \& Spillar, particularly because such
quantities will be measured by surveys occurring in the next decade.

In the process of studying this possibility, we have developed a
semi-analytic method for determining the abundance of dark halos as a
function of their circular velocity ($v_c^2 \equiv GM/r$) and redshift
that can provide results similar to those of N-body simulations in
considerably less computer time and without the effects of limited
resolution.  In essence, we begin with the abundances of galaxies as a
function of mass predicted by the modified Press-Schechter formalism of
Sheth \etal (1999), which provides an excellent fit to dark-matter
simulations.  We then determine the probability that a halo of given
mass has formation time such that it will virialize with the desired
circular velocity using the semi-analytic method of Lacey and Cole
(1994), which again is very well reproduced by simulations.  By
integrating over the distribution in mass we may then determine the
total abundance of objects with a given circular velocity at the epoch
of interest.  We describe this method in more detail in $\S$2, and the
prospects for our proposed test in $\S$3.

\section{Deriving the velocity function of dark halos}

A number of semi-analytic methods for calculating the veloctity
distribution of dark halos based upon the Press-Schechter
(\cite{ps74}) formalism have been proposed in the past.  Narayan \&
White (1988) used the assumption that halos have just virialized at
the epoch of observation to map the Press-Schechter distribution of
halos in mass into a distribution in velocity for a $\Omega=1$ Cold
Dark Matter (CDM) universe; Kochanek (1995) extended their analysis to
other common cosmological models.  However, the assumption that
objects observed are only just virializing is clearly inappropriate in
many cases, particularly for lower-mass halos; once objects virialize
and break away from the cosmological expansion, their circular
velocities evolve little if at all save in major mergers (Ghigna \etal
1999).  To account for this, Kitayama and Suto (1996a) proposed that
the distribution of X-ray temperatures of clusters ($T\propto v_c^2$)
be calculated by totalling the rate at which objects of the desired
temperature that survive without merging into a much more massive
object through the epoch of interest form, using the results of Lacey
\& Cole (1993).  As the authors of that paper noted, there remain some
ambiguities in their formulation, however.

We have attempted to improve upon these prior methods for calculating
the velocity distribution of dark matter halos, taking advantage of
the progress made over the past few years in extending the Press-Schechter formalism.  It has been repeatedly demonstrated that
standard Press-Schechter calculations overpredict the abundance of
low-mass halos compared to equivalent simulations; a number of
remedies have been suggested recently (e.g. Bond \& Myers 1996).  One
of the simplest to implement is that of Sheth \etal (1999), which
accounts for the average ellipticities of dark matter halos by making
the critical overdensity for their collapse mass-dependent.  The
resulting analytic expression for the number density of halos of given
mass (given by their equations 1 and 6) provides an excellent fit to
the results of a variety of cosmological simulations.
The results of Sheth \etal allow a prediction of halo abundances much
improved over the standard Press-Schechter formula with only minimal
complications to the calculation.  We have therefore used their equations
to determine the distribution of halos in mass at a given redshift in our work.

Because the density of the universe was greater at early times, objects that
virialized at high redshift are more compact than those that virialized
more recently, and therefore have a higher circular velocity for their mass.
Furthermore, once it has collapsed, the circular velocity of a halo
changes only minimally (Ghigna \etal 1999).  Therefore, in
determining the abundance of galaxy-scale halos with a given circular
velocity today, it is preferable to take into account the time of
formation of those halos in addition to their mass.  Fortunately, this
may be calculated; Lacey \& Cole (1993) derived a semi-analytic
formula for the distribution of formation times of halos with a given
mass that exist as virialized objects at a given redshift
(defining the formation time as the epoch when the object first had at
least half its final mass).  This formula provides an excellent fit to
the results of simulations (Lacey \& Cole 1994), even though it
assumes spherical collapse while the simulated halos could be
ellipsoidal.  We therefore have used their semi-analytic formula
(equation 2.19 of Lacey \& Cole 1994) to determine ${{\rm d}p/{\rm
d}t_f}$, the probability distribution of the epoch of collapse $t_f$
of halos observed to have mass $M$ at time $t_o$.
Note that the integral defining ${{\rm d}p/{\rm d}t_f}$ must be
evaluated with particular care where the integration variable $M_1
\approx M$.  However, an approximate analytic solution exists in that
region, and the integral may be safely performed over the remainder of
its range numerically.

In practice, we are interested in calculating the abundance of halos
at redshift $z$ that have circular velocity $v_c$; for each halo mass,
we therefore need only evaluate ${{\rm d}p/{\rm d}t_f}$ at that
formation epoch for which a halo of the given mass would virialize
with the desired circular velocity.  We may obtain that time from the
relationship between the circular velocity and the mass $M$, which is
most conveniently parameterized by using the length scale $r_0$
($h^{-1}$Mpc) $\equiv (3M/4\pi\rho_0)^{1/3}$, where $\rho_0$ is the
present-day mass density of the universe.  By combining the
definitions of circular velocity, of $r_0$, of $\Delta_{vir}$ (the
ratio of the mean density of a virialized halo to the critical density
of the universe when it virialized), and of the dimensionless
evolution of the Hubble parameter $E(z)$ we find (cf. Narayan \& White
1988, Kochanek 1995):
\begin{equation}
v_c= \Delta_{vir}^{1/6} \Omega_m^{1/3} {E(z)}^{1/3} {H_0 r_0 \over {\sqrt{2}~2^{1/3}}},
\end{equation}
where we have determined the circular velocity for a halo of mass
$M/2$, more appropriate given our definition of the formation time.  The
objection might be raised that we are using a formula for circular
velocity that implicitly assumes spherical geometry but are applying it to
potentially ellipsoidal dark matter halos; however, Lacey \& Cole
(1996) found that the velocity dispersions of dark matter halos
calculated using a singular isothermal sphere model (for which $\sigma
\equiv \sqrt{2} v_C$) matched the properties of N-body simulated halos
quite well.  In our calculations we have used the fitting formulae for
$\Delta_{vir}$ given by Bryan and Norman (1997).  

Thus, to calculate the comoving abundance of halos with a given circular
velocity at an epoch of interest, we simply calculate the distribution
of halos in mass on a 500-element grid, determine the probability that
those halos have the desired circular velocity, and then integrate
numerically over that grid.  This procedure takes $\sim5$ sec on a
modern workstation to calculate the abundance at a single circular
velocity and epoch, far faster than simulations could be run, and has
none of the limitations of resolution or sample size that affect even the
largest N-body simulations today.  Our technique seems to be similar
to one employed by Kitayama and Suto (1996b) to check their methods,
save for the use of the ellipsoidal-collapse mass function rather than
a traditional Press-Schechter calculation; however, details on the
latter are minimal.

Our procedures could certainly be improved upon.  For instance, the
connection between circular velocity, formation time, and mass could
be made using a more realistic model than a singular isothermal sphere
(\cite{nfw95}, Bullock \etal 1999).  Calculation of formation times
using an ellipsoidal-halo abundance paradigm rather than the
Press-Schechter method of Cole and Lacey might also yield modest
improvements.  Finally, we must note that the mass function of Sheth
\etal was based upon a fit to mass functions of halos identified in
N-body simulations using the ``friends-of-friends'' algorithm with
selection parameter $b=0.2$; in such a method, as in standard
Press-Schechter calculations, smaller subhaloes within a more massive
halo (e.g. galaxies within a cluster) are not counted separately in
the mass function.  Unless an improved method taking this into account
is used, we can only calculate the abundance of relatively {\bf
isolated} dark matter halos (i.e., those that are not part of a
larger, virialized group or cluster).  Such objects are readily
identified observationally in well-sampled redshift surveys, but this
is still a significant limitation.  Despite these caveats, we believe
that our technique for semi-analytic calculations provides a
substantial advance over methods previously used; all of them suffer
from similar flaws and have fewer advantages.  Our methods may also be
useful for a variety of cosmological calculations; for instance, the
circular velocity distribution of dark halos is required for
calculations of the abundance of strong gravitational lenses, and
hence for the limits placed on $\Omega_{\Lambda}$ thereby.

\section{The proposed test}

In order to study the evolution of the volume element of the universe,
we require some identifiable tracer whose number
density evolution is well known. The dark halos which contain galaxies are an
obvious candidate; the results of semi-analytic calculations (e.g.,
Sheth \etal 1999) and N-body simulations (e.g., \cite{virgo}) for the
behavior of dark matter are by now fairly well understood.  Of course,
directly observing the properties of dark halos at high redshift is
impossible; however, some of their parameters may be determined by
studying galaxies lying within them.  In particular, it is well-known
that the motion of material in the outer parts of typical galaxies is
dominated by the gravitational influence of dark matter (Fischer \etal
1999); the depth of the halo potential well may thus be directly
derived from observed linewidths or velocity dispersions of galaxies.
Unlike mass measurements, such observations are relatively insensitive
to distance from a galaxy's center (for instance, the rotation curves
of many spiral galaxies are observed to be fairly flat over a great
range of radii, as would be predicted for halos which are roughly
isothermal spheres).  Under modest assumptions (e.g. about their
radial profiles), the expected abundance of dark matter halos as a
function of observable linewidth may be determined from simulations or
from the semi-analytic method described in $\S$2.

Results of our semi-analytic calculations for the evolution of the
differential comoving abundance of isolated halos with circular
velocity 200 \kms\ in some commonly-used cosmological models are
depicted in Fig. 1.  A CDM-like power spectrum with shape parameter
$\Gamma=0.25$ and normalized to match the value of $\sigma_8$ derived
by Borgani \etal (1999) from X-ray observations of galaxy clusters was
employed in all cases.  As it is likely that selection effects will be
somewhat similar for galaxies at $z=0$ and $z\sim1$, we plot the ratio
of the abundance of galaxies at a given redshift to that at $z=0$.   The
abundance of halos at low circular velocities is smaller now than in
the recent past because of merging of smaller halos into new, larger
objects and because smaller halos lose their individual identities in
a Press-Schechter framework when they are included in a larger
virialized object such as a group or cluster.  Qualitatively, the
predictions of N-body simulations are similar (see, e.g., the cluster
simulations of Ghigna \etal 1999).
Our calculations demonstrate that the comoving abundance of halos with
fixed linewidth at $z\sim1$ relative to today is very insensitive to
the background cosmological model under reasonable assumptions about
the matter
power spectrum; e.g. for models with a low matter density $\Omega_m
=0.3$, the relative abundance is only $2\%$ different in a flat
model from an open one.  Thus, observations of the apparent numbers of
such objects per unit redshift per steradian in comparison to their
abundance today directly yields a measurement of the cosmological
volume element, and hence of fundamental cosmological parameters. 

A measurement of $dN/dzd\Omega$ will be quite feasible once the DEEP
Redshift Survey (\cite{deepwfsc}) is completed.  This survey, to be
begun in early 2001 and completed within the five years following,
will obtain spectra of $\sim60,000$ galaxies at redshift $z>0.7$ with
a resolution of $\sim80$ \kms\ FWHM, allowing measurements of
linewidths for perhaps a quarter to half of the galaxies in the
sample.  The number of spectra obtained will be great enough to test
for selection effects in redshift or linewidth by comparison to
calculations or simulations at the few percent level.  For instance,
smaller halos are predicted to be distributed in circular velocity
according to a power-law with a slope well-determined by both N-body
simulations and our techniques; deviation from such a distribution
would be a strong indicator of observational selection effects,
allowing us to correct for or avoid regions of parameter space
suffering from incompleteness.  Complementary information on the
linewidth distribution of galaxies at low redshift should be available
from the Sloan Digital Sky Survey (\cite{loveday98}), 2DF
(\cite{colless98}), or other local surveys, allowing accurate
normalization to the abundance of halos at $z=0$.


The determination of the volume element afforded by such measurements
can place constraints not only on the simplest models of the universe
which include only matter and a cosmological constant, but also on
so-called ``quintessence'' models (\cite{turnerwhite};
\cite{caldwell98}) in which the equation of state of a non-material
component, $P=w\rho$, can take arbitrary value between -1 and 1 ($w=0$
for matter, while $w=-1$ corresponds to a cosmological constant).  The
amount of comoving volume per unit redshift per steradian at $z\sim1$
depends strongly on $w$, as illustrated in Fig. 2.  The test we
propose in combination with a determination of $\Omega_m$ (e.g. from
velocity statistics of galaxies) should allow direct determination of
the equation of state of any dark energy present, as illustrated in
Figure 3.  In comparison, CMB measurements planned for the near future
and ground-based observations of the apparent magnitudes of Type Ia
supernovae can determine $w$ to $\sim10$\% at best (Hu \etal 1999);
similar accuracy might be achieved using the statistics of lensed
sources in the Sloan Digital Sky Survey (\cite{cooray99}).  Our
estimated constraints are similar in strength to those predicted for
the proposed SNAP satellite (Perlmutter et al. 1999), though less
sensitive to the present-day curvature and more sensitive in the
orthogonal direction (i.e., more complementary to CMB measurements).
The technique we propose is at least as effective as these even
without including complementary information from other tests.  As
Fig. 3 illustrates, since $\Omega_m$ will be increasingly constrained
from other experiments in the near future (including other analyses of
the DEEP dataset), the test we propose has the potential to determine
$w$ to $\sim 3\%$, even without other complementary measurements.
If the DEEP survey suffers from incompleteness, there
must be at least as many halos of a given linewidth as we observe;
thus, a lower limit on the volume element, and hence on $-w$, would be
provided.

The test we propose could also detect evolution of $w$, if any exists
and other cosmological parameters are well-determined.  It has
recently been proposed that in reasonable quintessence models $w$ may
vary with the current matter density as the universe evolves
(so-called ``tracker solutions''; cf. Steinhardt \etal 1998) or even
oscillate over time; such changes would be quite difficult to measure
via methods that study integrated quantities (e.g., the luminosity
distances of SNe Ia).  However, the test proposed here allows
measurement of a more localized quantity, the volume element; hence,
it may be able to measure changes in that quantity even on
relatively small scales.  By studying the apparent abundances of the
dark halos of galaxies, we may test the most recent cosmological
models -- or even those not yet envisioned -- with techniques having
their origin in the earliest days of extragalactic astronomy.


\vskip 36pt
We would like to thank Sandy Faber for her helpful suggestions and
James Bullock, Eric Gawiser, Dragan Huterer, James Robinson, Rob
Thacker, Mike Turner, Martin White, and the anonymous referee for
useful suggestions and discussions.  We would also like to thank
Zoltan Haiman and Jay Mohr for their assistance in including their
results in this paper.  This work was supported by NASA grant NAG
5-7833.



\clearpage
\vbox{%
\begin{center}
\leavevmode
\hbox{%
\epsfxsize=18cm
\epsffile{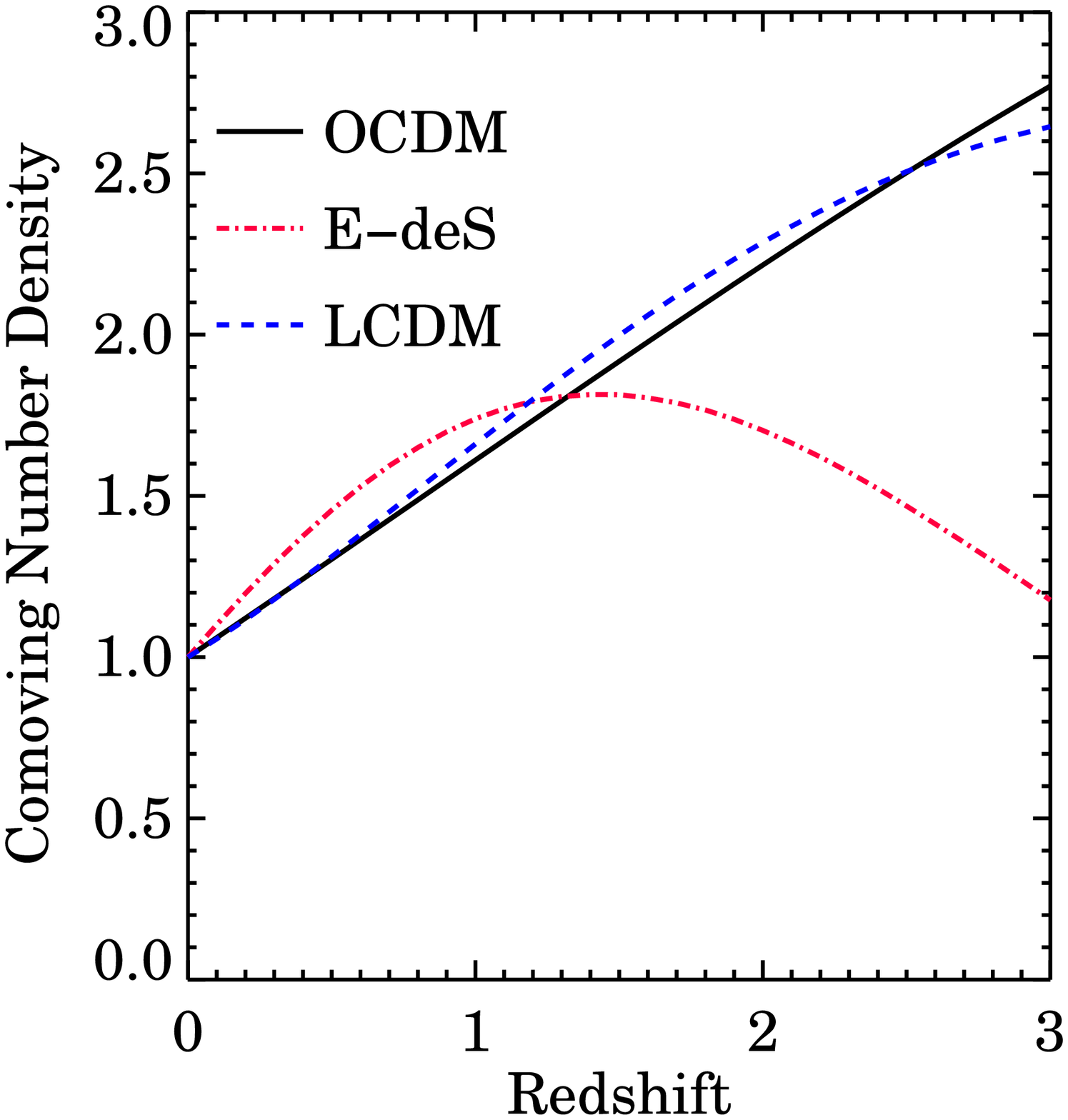}} \figcaption{\large The evolution of the
differential comoving number density of halos with circular velocity
200 \kms normalized to the value at $z=0$ .  We consider three common
cosmological models: an open Cold Dark Matter (CDM) model with
$\Omega=0.3$ today; an Einstein-De Sitter model with $\Omega=1$; and a
flat LCDM model having $\Omega_m=0.3$ today and a cosmological
constant.  Note the similarity of the predictions at $z\sim 1$.}
\end{center}}

\clearpage

\begin{center}
\leavevmode
\hbox{%
\epsfxsize=18cm
\epsffile{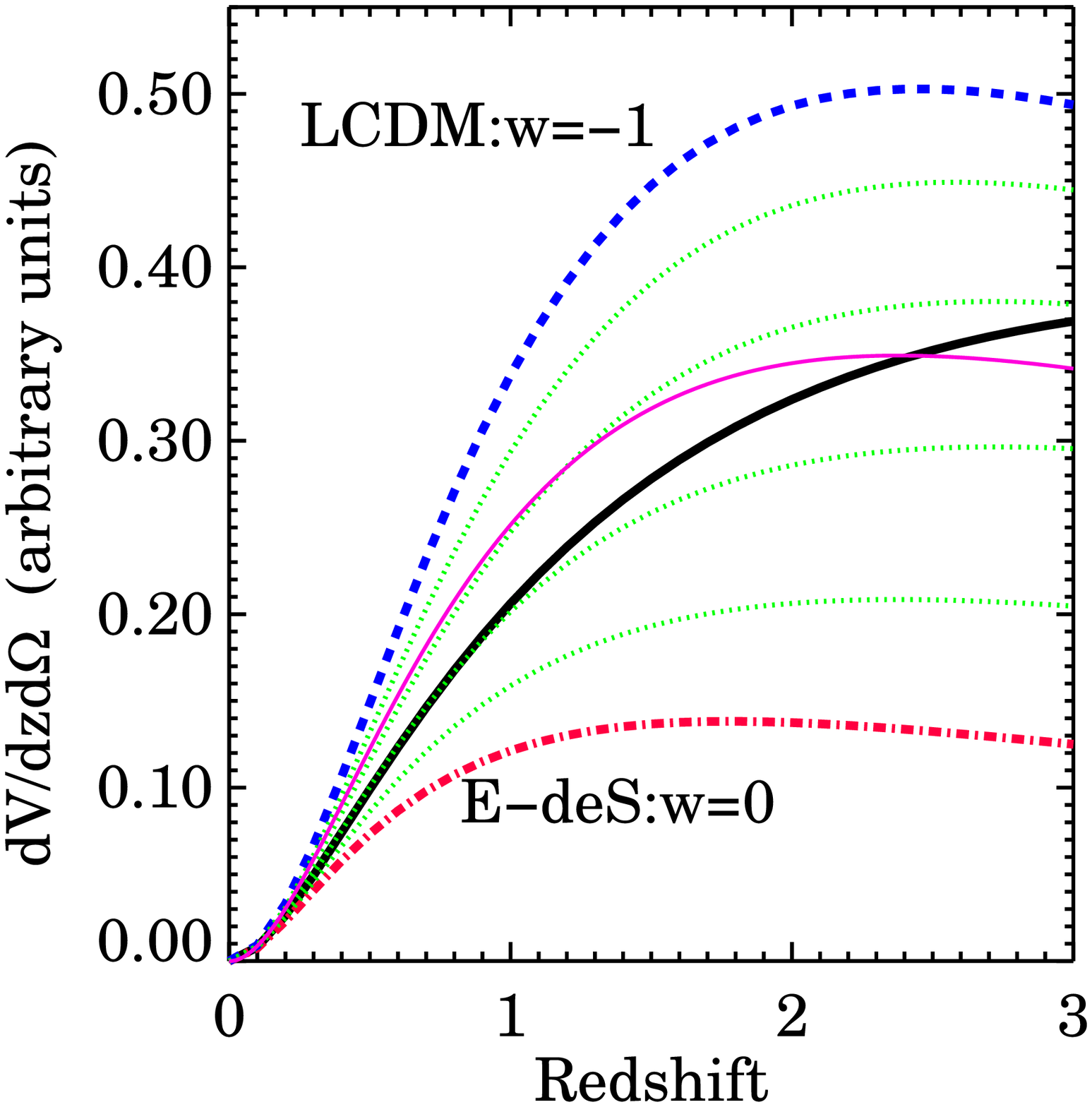}}
\figcaption{\large The comoving volume per unit redshift per
steradian, $dV/dzd\Omega$, in a variety of cosmological models
(cf. Peebles 1993).  The thickest curves indicate the models
considered in Fig. 1; the thick solid black curve depicts the volume
element in a $\Omega_m=0.3$ open universe, the red dot-dashed curve
shows an Einstein-de Sitter model, and the blue dashed curve depicts
an LCDM model.  The green, light dotted curves show the evolution of
the volume element in standard quintessence models with
$\Omega_m=0.3$, $\Omega_Q=0.7$, and constant $w=-0.2$, -0.4, -0.6, and
-0.8 (listed from lowest-volume to highest).  Finally, the thin, solid
purple line depicts a model in which $w$ is proportional to the
expansion parameter, $w=-0.8a$.}
\end{center} \vskip 12pt

\clearpage

\begin{center}
\leavevmode
\hbox{%
\epsfxsize=15.5cm
\epsffile{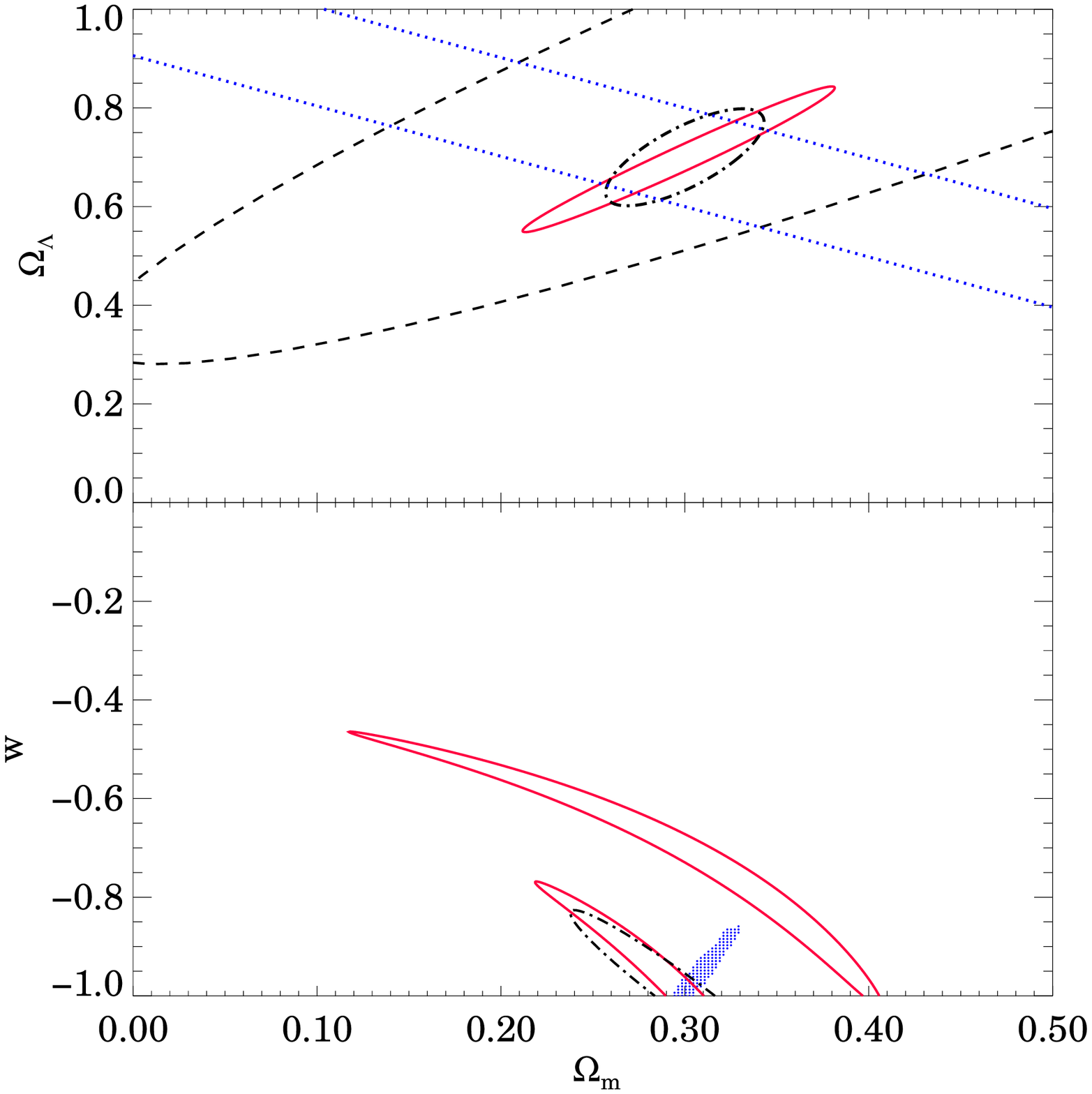}}
\figcaption{\large Predicted $95\%$ confidence level contours for
application of the d$N/$d$z$ test to the DEEP Redshift Survey (solid
red curves), along with other existing or predicted constraints.
These have been obtained by conservatively assuming that the DEEP
Redshift Survey will obtain useful linewidth information on 10,000
galaxies distributed as $(1+z)^{-2}$ among 8 bins between z=0.7 and
1.5 (i.e., that the great majority of observed galaxies are excluded
from the test because of potential systematic errors or
incompleteness).  The top panel depicts confidence contours in the
$\Omega_m-\Omega_{\Lambda}$ plane (i.e., under the assumption that
$w=-1$) about the point ($\Omega_m=0.3$, $\Omega_{\Lambda}=0.7$).
Overplotted are the current 68\% confidence SNeIa constraint from
Perlmutter $et al.$ l 1999 (black, dashed line), the target 95\% confidence
interval for the proposed SNAP satellite (dot-dashed black curve;
taken from figures on the SNAPSAT website at http://snap.lbl.gov), and
a sample CMB constraint (Melchiorri $et al.$ 1999) with errors reduced to
simulate a determination of $\Omega_{curvature}$ to $\pm 0.1$,
representative of what may be expected from experiments performed
before DEEP is completed such as BOOMERANG and MAP.   The solid red curves in the bottom panel show the
predicted 95\% confidence contour from the DEEP Redshift Survey in the
$\Omega_m-w$ plane (under the assumption of a zero-curvature universe)
about the points ($\Omega_m=0.3$, $w=-0.7$) and ($\Omega_m=0.3$,
$w=-1$).  Also plotted are target 95\% confidence constraints for the
latter model from SNAPSAT (black dot-dashed curve) and from
observations of cluster abundances by the
proposed COSMEX satellite (blue dotted region; cf. Haiman et
al. 2000).  The techniques involved (and thus, almost certainly, any
systematic errors) differ greatly among these methods, making each a
valuable check on the others.}
\end{center} \vskip 12pt

\end{document}